\documentclass[twocolumn,superscriptaddress,letter]{revtex4}%
\usepackage{amssymb}
\usepackage{color}
\usepackage{graphicx}
\usepackage{dcolumn}
\usepackage{bm}
\usepackage[header,title,page,titletoc]{appendix}
\usepackage{amsmath}
\usepackage{amsfonts}%
\providecommand{\U}[1]{\protect \rule{.1in}{.1in}}

\begin{document}
\title{Anomalous Spontaneous Symmetry Breaking in non-Hermitian Systems with
Biorthogonal Z$_{\emph{2}}$-symmetry}
\author{Meng-Lei Yang}
\affiliation{Center for Advanced Quantum Studies, Department of Physics, Beijing Normal
University, Beijing 100875, China}
\author{Heng Wang}
\affiliation{Center for Advanced Quantum Studies, Department of Physics, Beijing Normal
University, Beijing 100875, China}
\author{Cui-Xian Guo}
\affiliation{Center for Advanced Quantum Studies, Department of Physics, Beijing Normal
University, Beijing 100875, China}
\author{Xiao-Ran Wang}
\affiliation{Center for Advanced Quantum Studies, Department of Physics, Beijing Normal
University, Beijing 100875, China}
\author{Gaoyong Sun}
\thanks{Corresponding author}
\email{gysun@nuaa.edu.cn}
\affiliation{College of Science, Nanjing University of Aeronautics and Astronautics,
Nanjing, 211106, China}
\author{Su-Peng Kou}
\thanks{Corresponding author}
\email{spkou@bnu.edu.cn}
\affiliation{Center for Advanced Quantum Studies, Department of Physics, Beijing Normal
University, Beijing 100875, China}

\begin{abstract}
Landau's spontaneous symmetry breaking theory is a fundamental theory that
describes the collective behaviors in many-body systems. It was well known
that for usual spontaneous symmetry breaking in Hermitian systems, the
order-disorder phase transition with gap closing and spontaneous symmetry
breaking occur at the same critical point. In this paper, we generalized the
Landau's spontaneous symmetry breaking theory to the cases in non-Hermitian
(NH) many-body systems with biorthogonal \textrm{Z}$_{\mathit{2}}$ symmetry and
tried to discover certain universal features. We were surprised to find that
the effect of the NH terms splits the spontaneous biorthogonal \textrm{Z}%
$_{\mathit{2}}$ symmetry breaking from a (biorthogonal) order-disorder phase
transition with gap closing. The sudden change of similarity for two
degenerate ground states indicates a new type of quantum phase transition
without gap closing accompanied by spontaneous biorthogonal \textrm{Z}%
$_{\mathit{2}}$ symmetry breaking. We will take the NH transverse Ising model
as an example to investigate the anomalous spontaneous symmetry breaking. The
numerical results were consistent with the theoretical predictions.

\end{abstract}

\maketitle

Landau's spontaneous symmetry breaking theory is a fundamental theory and
plays an important role in modern particle physics and condensed matter
physics. For a process with \emph{spontaneous symmetry breaking}, the model of
the system obeys \emph{symmetries}, but the ground state (or vacuum) does not
exhibit the same symmetry. It is perturbations that select one from the
degenerate ground states. A lot of universal features become physics
consequences of the spontaneous symmetry breaking, including the
order-disorder phase transition, the universal critical phenomenon, the
symmetry-protected degeneracy of ground states, ...

On the other hand, because non-Hermitian (NH) systems show quite different
properties with their Hermitian counterparts, it attracts massive researches
from different fields in recent years\cite{Bender
98,Rudner2009,Esaki2011,Hu2011,Liang2013,Zhu2014,Lee2016,San2016,Leykam2017,Shen2018,Lieu2018,
Xiong2018,Kawabata2018,Gong2018,Yao2018,YaoWang2018,Yin2018,Kunst2018,KawabataUeda2018,Alvarez2018,
Jiang2018,Ghatak2019,Avila2019,Jin2019,Lee2019,Liu2019,38-1,38,chen-class2019,Edvardsson2019,
Herviou2019,Yokomizo2019,zhouBin2019,Kunst2019,Deng2019,SongWang2019,xi2019,Longhi2019,chen-edge2019,WangC2019,XiaoR2020,GuoC2020}%
. For a special type of NH models, there may also exist real spectra. For
parity-time ($\mathcal{PT}$) symmetric models\cite{Bender 98} that is
invariant under the combined action of the $\mathcal{P}$ and $\mathcal{T}$
operations, the energy spectra are real in $\mathcal{PT}$-symmetric phase. In
$\mathcal{PT}$-symmetric phase, the system always obeys similarities, i.e., it
could be deformed to a Hermitian system under a NH similarity transformation
(ST). For example, for certain NH transverse Ising models with imaginary
external field\cite{Olalla09,Deguchi09,James13,Alexander13,James14,ZSong14,ZSong16,Simon19,Yoshihiro20,ZSong20,Liang19}, one can apply a certain ST and map the original NH transverse
Ising model to a Hermitian model with the same energy
spectra.

In this paper, we ask the following questions, "\emph{How spontaneous symmetry
occur in many-body NH systems?}" and "\emph{Do there exist new universal
features for the spontaneous symmetry breakings in NH systems?}" Motivated by
above questions, we investigate a special class of NH systems with the
non-unitary \textrm{Z}$_{\mathit{2}}$ symmetry (or the so-called biorthogonal
\textrm{Z}$_{\mathit{2}}$-symmetry) and try to develop the theory for
\emph{non-Hermitian spontaneous symmetry breaking} (NHSSB). The universal
features of NHSSB for this class of NH systems are explored.\

\textbf{Biorthogonal Z}$_{\emph{2}}$\textbf{-symmetry in NH systems: }Firstly,
we discuss the global \textrm{Z}$_{\mathit{2}}$-symmetry in Hermitian systems.
For a Hermitian system with \textrm{Z}$_{\mathit{2}}$-symmetry, the
Hamiltonian $\mathrm{\hat{H}}_{Z_{2}}$ is invariant under the group operation
$g\in \mathrm{Z}_{\mathit{2}}$, i.e., $\hat{U}(g)\mathrm{\hat{H}}_{Z_{2}}%
\hat{U}^{-1}(g)=\mathrm{\hat{H}}_{Z_{2}}\quad$where $\hat{U}(g)$ is the
unitary operator (with $\det(\hat{U}(g))\equiv1$) representing the operation
$g$ on the Hilbert space. In general, the local degrees of freedom that has a
nonvanishing ground-state expectation value for a system with \textrm{Z}%
$_{\mathit{2}}$-symmetry can be phenomenologically described by a 1/2
(pseudo)-spin operator $\tau_{i}^{z}$, i.e., $(%
\begin{array}
[c]{c}%
1\\
0
\end{array}
)$ or $(%
\begin{array}
[c]{c}%
0\\
1
\end{array}
)$. The group operation is $\hat{U}(g)=%
{\displaystyle \prod \limits_{i}}
\tau_{i}^{\theta}$ where $\tau_{i}^{\theta}=\tau_{i}^{x}\cos \theta+\tau
_{i}^{y}\sin \theta$ ($\theta$ is a real number). Under this group
transformation, we have $\hat{U}(g)\tau_{i}^{z}\hat{U}^{-1}(g)=-\tau_{i}^{z}$.

Next, we introduce the NH generalization for global \textrm{Z}$_{\mathit{2}}%
$-symmetry in a NH system -- \emph{biorthogonal }\textrm{Z}$_{\mathit{2}}%
$\emph{-symmetry}.\

\textit{Definition 1 -- Biorthogonal }\textrm{Z}$_{\mathit{2}}$%
\textit{-symmetry: For a NH Hamiltonian }$\mathrm{\hat{H}}_{\tilde{Z}_{2}}%
$\textit{ (with }$\mathrm{\hat{H}}_{\tilde{Z}_{2}}^{\dagger}\neq
\mathrm{\hat{H}}_{\tilde{Z}_{2}}$\textit{), there exists a global non-unitary
}\textrm{Z}$_{\mathit{2}}$\textit{ symmetry, i.e., }$\tilde{g}\in \tilde{Z}%
_{2}:$\textit{ }$\tilde{U}(\tilde{g})\mathrm{\hat{H}}_{\tilde{Z}_{2}}\tilde
{U}^{-1}(\tilde{g})=\mathrm{\hat{H}}_{\tilde{Z}_{2}}$\textit{ (or }$\tilde
{U}(\tilde{g}^{-1})\mathrm{\hat{H}}_{\tilde{Z}_{2}}^{\dagger}\tilde{U}%
^{-1}(\tilde{g}^{-1})$\textit{). Here, }$\hat{U}(\tilde{g})$\textit{ (or
}$\hat{U}^{\dagger}(\tilde{g})$) is\textit{ the non-unitary operator (with
}$\det(\hat{U}(\tilde{g}))\neq1$\textit{ or }$\det(\hat{U}^{\dagger}(\tilde
{g}))\neq1$)\textit{ representing }$\tilde{g}$\textit{ (or }$\tilde{g}^{-1}%
$)\textit{ on the Hilbert space} \textit{that obeys }$\tilde{U}(\tilde
{g})\cdot \tilde{U}(\tilde{g}^{-1})=1$\textit{. }For a (pseudo)-spin operator
with nonvanishing ground-state expectation value $\tau_{i}^{z},$ we have
$\hat{U}(\tilde{g})\tau_{i}^{z}\hat{U}^{-1}(\tilde{g})=-\tau_{i}^{z}$ (or
$\hat{U}(\tilde{g}^{-1})\tau_{i}^{z}\hat{U}^{-1}(\tilde{g}^{-1})=-\tau_{i}%
^{z}$).

In general, except for a possible unitary transformation the group operation
for biorthogonal \textrm{Z}$_{\mathit{2}}$-symmetry can be transformed into a
unitary one $\hat{U}(g)$ for usual global \textrm{Z}$_{\mathit{2}}$-symmetry
by a similarity transformation, i.e., $\hat{U}(\tilde{g})=\mathcal{S}^{-1}%
\hat{U}(g)\mathcal{S}$. That means the NH system with biorthogonal
\textrm{Z}$_{\mathit{2}}$-symmetry ($\tilde{U}(\tilde{g})\mathrm{\hat{H}%
}_{\tilde{Z}_{2}}\tilde{U}^{-1}(\tilde{g})=\mathrm{\hat{H}}_{\tilde{Z}_{2}}$)
\emph{obeys} global similarities (STs), $\mathcal{S}^{-1}\mathrm{\hat{H}%
}_{\tilde{Z}_{2}}\mathcal{S}=\mathrm{\hat{H}}_{Z_{2}}$ where $\mathcal{S}$ is
the operator for the (NH) ST and $\mathrm{\hat{H}}_{Z_{2}}$ is a (Hermitian or
NH) Hamiltonian obeying global \textrm{Z}$_{\mathit{2}}$-symmetry, i.e.
$\hat{U}(g)\mathrm{\hat{H}}_{Z_{2}}\hat{U}^{-1}(g)=\mathrm{\hat{H}}_{Z_{2}}.$
If the local degrees of freedom that has a nonvanishing ground-state
expectation value is denoted by a pseudo-spin operator $\tau_{i}^{z},$ we can
phenomenologically derive $\mathcal{S}$ to be $\mathcal{S}(\beta)=%
{\displaystyle \prod \limits_{i}}
S_{i}(\beta)$ with $S_{i}(\beta)=(%
\begin{array}
[c]{cc}%
{1} & 0\\
0 & {\mathrm{e}^{-\beta}}%
\end{array}
).$ Here, $\beta$ denotes the non-Hermiticity. In the following parts, in
order to be more obvious, we denote the non-unitary group operator $\tilde{g}$
by $g^{\beta}$ and $\mathcal{S}$ by $\mathcal{S}(\beta).$ And when $\beta=0$,
the NH model turns into a Hermitian one.

\begin{figure}[ptb]
\includegraphics[clip,width=0.5\textwidth]{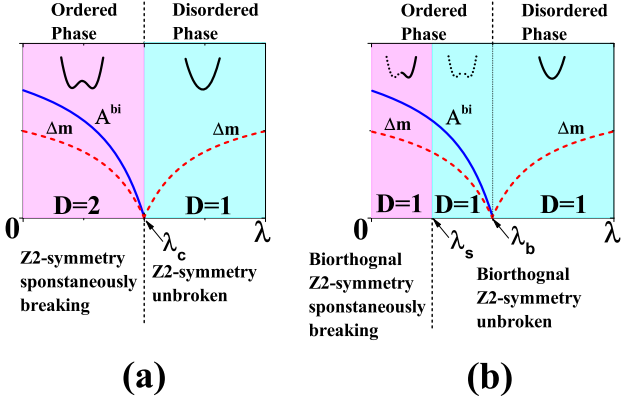}\caption{(Color online)
(a) The scheme of spontaneous \textrm{Z}$_{\mathit{2}}$-symmetry breaking for
Hermitian systems. In general, there are two phases. At $\lambda=\lambda_{c},$
the order-disorder phase transition occurs together with spontaneous
\textrm{Z}$_{\mathit{2}}$-symmetry breaking; (b) The scheme of anomalous
spontaneous \textrm{Z}$_{\mathit{2}}$-symmetry breaking for non-Hermitian
systems. In general, there are three phases. The spontaneous biorthogonal
\textrm{Z}$_{\mathit{2}}$-symmetry breaking becomes a quantum phase transition
at $\lambda=\lambda_{s}$ without gap closing that separates with the
biorthogonal order-disorder phase transition with gap closing at
$\lambda=\lambda_{c}$. For a double-well potential with two hidden wells,
there is only one ground state in the double-well system; For double-well
potential with a hidden well denoted by dotted curve, there is no quantum
states in it. The two cases are both physics consequences of the EPs with NH
coalescence for the two degenerate ground states.}%
\end{figure}

\textbf{Universal features for non-Hermitian spontaneous symmetry
breaking:}\textit{ }Firstly, as shown in Fig.1(a), we summarize the universal
features for usual spontaneous \textrm{Z}$_{\mathit{2}}$-symmetry breaking in
Hermitian systems: 1) The energy gap for bulk states is closed ($\Delta m=0$)
at $\lambda=\lambda_{c}$ (with $\lambda$ is a tunable parameter). For example,
$\lambda$ is the strength of a transverse field in transverse Ising model; 2)
A quantum phase transition (QPT) occurs at $\lambda=\lambda_{c}$ from ordered
phase with $\left \langle \mathrm{vac}|\hat{A}|\mathrm{vac}\right \rangle
=A_{0}\neq0$ to disordered phase $\left \langle \mathrm{vac}|\hat
{A}|\mathrm{vac}\right \rangle =A_{0}=0.$ Here, $\left \vert \mathrm{vac}%
\right \rangle $ denotes ground state and $\hat{A}$ is an operator with a
nonvanishing ground-state expectation value which changes sign under
$\mathrm{Z}_{\mathit{2}}$ group, $\hat{U}(g)\hat{A}\hat{U}^{-1}(g)=-\hat{A}$.
In general, $\hat{A}$ is denoted by a pseudo-spin operator $\tau_{i}^{z}$; 3)
The ground state degeneracy $D$ changes suddenly from $2$ in ordered phase to
$1$ in disordered phase.

Therefore, as shown in Fig.1(a), for usual spontaneous \textrm{Z}%
$_{\mathit{2}}$-symmetry breaking in Hermitian systems ($\hat{U}%
(g)\mathrm{\hat{H}}_{Z_{2}}\hat{U}^{-1}(g)=\mathrm{\hat{H}}_{Z_{2}}$), there
are \emph{two} phases: ordered phase and disordered phase. In the ordered
phase ($A_{0}\neq0$), the ground states have two-fold degeneracy. Under
perturbations that select one from the degenerate ground states (DGSs),
spontaneous \textrm{Z}$_{\mathit{2}}$-symmetry breaking occurs, i.e., $\hat
{U}(g)\left \vert \mathrm{vac}\right \rangle \neq \left \vert \mathrm{vac}%
\right \rangle $; in disordered phase ($A_{0}=0$), \textrm{Z}$_{\mathit{2}}%
$-symmetry is unbroken, i.e., $\hat{U}(g)\left \vert \mathrm{vac}\right \rangle
=\left \vert \mathrm{vac}\right \rangle $.

To illustrate the universal features for the NHSSB in NH systems with
biorthogonal \textrm{Z}$_{\mathit{2}}$-symmetry, we introduce a concept --
\emph{spontaneous biorthogonal }\textrm{Z}$_{\mathit{2}}$\emph{-symmetry
breaking}, together with two theorems (the detailed proof is given in
supplementary materials):

\textit{Definition 2 -- Spontaneous biorthogonal }\textrm{Z}$_{\mathit{2}}%
$\textit{-symmetry breaking: For non-Hermitian systems with biorthogonal
}\textrm{Z}$_{\mathit{2}}$-\textit{symmetry} \textit{(}$\hat{U}(\tilde
{g})\mathrm{\hat{H}}_{\tilde{Z}_{2}}\hat{U}^{-1}(\tilde{g})=\mathrm{\hat{H}%
}_{\tilde{Z}_{2}}$\textit{), the ground states (or vacuum) shows the same
symmetry }$\hat{U}(\tilde{g})\left \vert \mathrm{\overline{vac}}(\beta
)\right \rangle =\pm \left \vert \mathrm{\overline{vac}}(\beta)\right \rangle
.$\textit{ However, with an additional perturbation }$\mathrm{\hat{H}}%
_{\tilde{Z}_{2}}\rightarrow \mathrm{\hat{H}}_{\tilde{Z}_{2}}+\delta
\mathrm{\hat{H},}$\textit{ the ground states (or vacuum) does not exhibit the
original symmetry, i.e., }$\hat{U}(\tilde{g})\left \vert \mathrm{\overline
{vac}}(\beta)\right \rangle \neq \pm \left \vert \mathrm{\overline{vac}}%
(\beta)\right \rangle .$

\textit{Theorem 1: Near the biorthogonal order-disorder QPT for the NH system
described by }$\mathrm{\hat{H}}_{\tilde{Z}_{2}}$\textit{ with biorthogonal
Z}$_{\emph{2}}$\textit{-symmetry, under the conditions }$\mathrm{\hat{H}%
}_{Z_{2}}=\mathcal{S}^{-1}(\beta)\mathrm{\hat{H}}_{\tilde{Z}_{2}}%
\mathcal{S}(\beta)$ and $\mathrm{\hat{H}}_{Z_{2}}=\mathrm{\hat{H}}_{Z_{2}%
}^{\dagger}$\textit{, the universal critical phenomenon for biorthogonal order
parameter is the same to that of the Hermitian model }$\mathrm{\hat{H}}%
_{Z_{2}}.$ Here $\mathcal{S}(\beta)$ is a global ST\textit{.}\emph{\ }A
\emph{biorthogonal order parameter} is defined by calculating the expectation
value for the DGSs in the biorthogonal set $\left \vert \mathrm{\overline{vac}%
}^{\mathrm{R}}(\beta)\right \rangle $ and $\left \vert \mathrm{\overline{vac}%
}^{\mathrm{L}}(\beta)\right \rangle $\cite{Mostafazadeh02}, i.e., $\left \langle
\mathrm{\overline{vac}}^{\mathrm{L}}(\beta)|\tilde{A}|\mathrm{\overline{vac}%
}^{\mathrm{R}}(\beta)\right \rangle =A^{\mathrm{bi}}$ where $\hat{A}$ is an
operator with a nonvanishing ground-state expectation value which transforms
non-trivially under biorthogonal $\mathrm{Z}_{\mathit{2}}$ group, $\tilde
{U}(g^{\beta})\tilde{A}\tilde{U}^{-1}(g^{\beta})=-\tilde{A}$.

\textit{Theorem 2}: \textit{For the NH system in biorthogonal ordered phase,
with increasing the non-Hermiticity }$\beta$ \textit{the biorthogonal
Z}$_{\emph{2}}$-symmetry\textit{ could be spontaneously broken that is
accompanied by the sudden change of state-similarity }$\left \vert
\langle \mathrm{\overline{vac}}_{-}(\beta)|\mathrm{\overline{vac}}_{+}%
(\beta)\rangle \right \vert $\textit{ for the two DGSs }$|\mathrm{\overline
{vac}}_{\pm}(\beta)\rangle$\textit{. Here, }$|\mathrm{\overline{vac}}_{\pm
}(\beta)\rangle$\textit{ is satisfied self-normalization condition, i.e.,
}$|\langle \mathrm{\overline{vac}}_{+}(\beta)|\mathrm{\overline{vac}}_{+}%
(\beta)\rangle|\equiv1,$ $|\langle \mathrm{\overline{vac}}_{-}(\beta
)|\mathrm{\overline{vac}}_{-}(\beta)\rangle|\equiv1$\textit{) for the two DGSs
in biorthogonal order. }

As shown in Fig.1(b), in general, for NH systems with biorthogonal
\textrm{Z}$_{\mathit{2}}$-symmetry, there are \emph{three} phases: 1) a
biorthogonal order (BO) with $A^{\mathrm{bi}}\neq0$, in which the biorthogonal
\textrm{Z}$_{\mathit{2}}$-symmetry are spontaneously broken simultaneously,
i.e., $\tilde{U}(g^{\beta})\left \vert \mathrm{\overline{vac}}(\beta
)\right \rangle \neq \left \vert \mathrm{\overline{vac}}(\beta)\right \rangle $.
In the inset in Fig.1(b), we can effectively use a figure of double-well
potential with a hidden well to represent this ordered phase (there is no
quantum states in the hidden well); 2) a BO with $A^{\mathrm{bi}}\neq0$, in
which the biorthogonal \textrm{Z}$_{\mathit{2}}$-symmetry are unbroken, i.e.,
$\tilde{U}(g^{\beta})\left \vert \mathrm{\overline{vac}}(\beta)\right \rangle
=\pm \left \vert \mathrm{\overline{vac}}(\beta)\right \rangle .$ In the inset in
Fig.1(b), we can effectively use a figure of double-well potential with two
hidden wells to represent the this ordered phase (there is only one ground
state in the double-well system); 3) a disordered phase with $A^{\mathrm{bi}%
}=0$, in which the biorthogonal \textrm{Z}$_{\mathit{2}}$-symmetry are
unbroken, i.e., $\tilde{U}(g^{\beta})\left \vert \mathrm{\overline{vac}}%
(\beta)\right \rangle =\left \vert \mathrm{\overline{vac}}(\beta)\right \rangle
.$ In the inset in Fig.1(b), we can effectively use a figure of single-well
potential to represent the disordered phase.

In Fig.1(b), the biorthogonal order-disorder QPT with gap closing occurs at
$\lambda=\lambda_{c}$. However, another QPT without gap closing occur at
$\lambda=\lambda_{s}$, at which the biorthogonal \textrm{Z}$_{\mathit{2}}%
$-symmetry are spontaneously breaking. We can say it is the effect of NH terms
that splits the spontaneous biorthogonal \textrm{Z}$_{\mathit{2}}$-symmetry
breaking (at $\lambda=\lambda_{s}$) from the biorthogonal order-disorder QPT
(at $\lambda=\lambda_{c}$) with $\lambda_{s}\neq \lambda_{c}$.

\textbf{Example -- One dimensional transverse Ising model with biorthogonal
}\textrm{Z}$_{\mathit{2}}$\textbf{ symmetry:}\textit{ }We use a transverse
Ising (TI) model\cite{Ovchinnikov03,Pfeuty,Elliott,Suzuki13} with biorthogonal \textrm{Z}$_{\mathit{2}}$ symmetry as an
example to show the universal features of NHSSB for NH
systems.

The Hamiltonian of one dimensional (1D) TI model with biorthogonal Z$_{2}$
symmetry is given by
\begin{equation}
\mathrm{\hat{H}}_{\mathrm{NTI}}^{\beta}=\sum_{i}(-J\sigma_{i}^{x}\sigma
_{i+1}^{x}+h^{y}\sigma_{i}^{y}+ih^{z}\sigma_{i}^{z})
\end{equation}
where $J>0$ is ferromagnetic Ising coupling constant between two nearest
neighbor spins and $h^{y}$ is the strength of a real transverse field along
\textrm{y}-axis, $h^{z}$ is the strength of a imaginary transverse field along
\textrm{z}-axis. In this paper, the coupling parameter $J$ is set to be unit,
$J\equiv1$.

The group element of the biorthogonal \textrm{Z}$_{\mathit{2}}$-symmetry is
defined by a non-unitary operator $\tilde{U}(g^{\beta})=\mathcal{S}%
(\beta)[\prod_{i}(\sigma_{i}^{y})]\mathcal{S}(\beta)^{-1},$ where $\det
[\tilde{U}(g^{\beta})]\neq1$ (or $\det(\tilde{U}^{\dagger}(g^{\beta}))\neq1$)
and $\tilde{U}(g^{\beta})\tilde{U}(g^{-\beta})=1$. Here $\mathcal{S}(\beta)=%
{\displaystyle \prod \limits_{i}}
S_{i}(\beta)$ is the operator of a global NH ST on spin system. The similar
transformation $S_{i}(\beta)$ is defined as $S_{i}(\beta)=\frac{1}{2}%
\begin{pmatrix}
1+e^{-\beta} & 1-e^{-\beta}\\
1-e^{-\beta} & 1+e^{-\beta}%
\end{pmatrix}
$ and the non-Hermiticity is $\beta=\ln(\frac{h^{y}+h^{z}}{h^{y}-h^{z}})$.
According to $\tilde{U}^{-1}(g^{\beta})\cdot \mathrm{\hat{H}}_{\mathrm{NTI}%
}^{\beta}\cdot \tilde{U}(g^{\beta})=\mathrm{\hat{H}}_{\mathrm{NTI}}^{\beta
}\quad$(or $\tilde{U}^{-1}(g^{-\beta})\cdot(\mathrm{\hat{H}}_{\mathrm{NTI}%
}^{\beta})^{\dagger}\cdot \tilde{U}(g^{-\beta})=(\mathrm{\hat{H}}%
_{\mathrm{NTI}}^{\beta})^{\dagger}$) for $\tilde{U}(g^{\beta})=\mathcal{S}%
(\beta)[\prod_{i}(\sigma_{i}^{y})]\mathcal{S}(\beta)^{-1}$, $\mathrm{\hat{H}%
}_{\mathrm{NTI}}^{\beta}$ shows\emph{ biorthogonal Z}$_{\emph{2}}%
$\emph{-symmetry} and global similarity.

Because the original TI\ model can be written into $\mathrm{\hat{H}%
}_{\mathrm{NTI}}^{\beta}=\sum_{i}(-J\sigma_{i}^{x}\sigma_{i+1}^{x}%
+h(\sigma_{i}^{y})^{\beta}),$ where $(\sigma_{i}^{y})^{\beta}=S_{i}%
(\beta)\sigma_{i}^{y}S_{i}(\beta)^{-1}=\cosh(\beta)\sigma_{i}^{y}+i\sinh
(\beta)\sigma_{i}^{z},$ and $h=\sqrt{\left \vert h^{y}\right \vert
^{2}-\left \vert h^{z}\right \vert ^{2}},$ under a global inverse ST,
$\mathrm{\hat{H}}_{\mathrm{NTI}}^{\beta}$ is deformed into a Hermitian one,
i.e.,
\begin{equation}
\mathrm{\hat{H}}_{\mathrm{NTI}}^{\beta=0}=\mathcal{S}^{-1}(\beta
)\mathrm{\hat{H}}_{\mathrm{NTI}}^{\beta}\mathcal{S}(\beta)=\sum_{i}%
(-J\sigma_{i}^{x}\sigma_{i+1}^{x}+h\sigma_{i}^{y}).
\end{equation}
A $\mathcal{PT}$ spontaneous symmetry breaking occurs at $\left \vert
h^{y}\right \vert =\left \vert h^{z}\right \vert $\cite{com}. For the case of
$\left \vert h^{y}\right \vert >\left \vert h^{z}\right \vert ,$ the energy
spectra for excitations are all real; For the case of $\left \vert
h^{y}\right \vert <\left \vert h^{z}\right \vert $, the energy spectra for the
excitations become complex.

Firstly, we study the QPT at gap closing. In this part, we focus on the
$\mathcal{PT}$ symmetric phase, $\left \vert h^{y}\right \vert >\left \vert
h^{z}\right \vert $. Under NH (inverse) ST, the energy levels $E_{n}(\beta)$ of
$\mathrm{\hat{H}}_{\mathrm{NTI}}^{\beta}$ (or $(\mathrm{\hat{H}}%
_{\mathrm{NTI}}^{\beta})^{\dagger}$) are same to those of the Hermitian model
$E_{n}(\beta=0)$ of $\mathrm{\hat{H}}_{\mathrm{NTI}}^{\beta=0}$, i.e.,
$E_{n}(\beta)=E_{n}(-\beta)=E_{n}(\beta=0).$\emph{\ }As a result, the QPT with
the gap closing for $\mathrm{\hat{H}}_{\mathrm{NTI}}^{\beta}$ is same to that
for $\mathrm{\hat{H}}_{\mathrm{NTI}}^{\beta=0}$ that is obtained as
$J=\left \vert h\right \vert $\cite{com1}. See Fig.2(a) from exact diagonal
numerical calculation for 1D NH TI model with $N=16$, in which the dotted red
lines come from theoretical prediction.

Next, we study the biorthogonal order-disorder phase transition. The
biorthogonal order parameter is defined by the expectation value in the ground
states $\left \vert \mathrm{\overline{vac}}^{\mathrm{L}}(\beta)\right \rangle $
and $\left \vert \mathrm{\overline{vac}}^{\mathrm{R}}(\beta)\right \rangle $,
i.e., $\frac{1}{N}%
{\displaystyle \sum \limits_{i}}
\left \langle \mathrm{\overline{vac}}^{\mathrm{L}}(\beta)|\sigma_{i}%
^{x}|\mathrm{\overline{vac}}^{\mathrm{R}}(\beta)\right \rangle =A^{\mathrm{bi}%
}$. In the region of $J>\left \vert h\right \vert $, $A^{\mathrm{bi}}\neq0,$
there exists BO; In the region of $J<\left \vert h\right \vert $,
$A^{\mathrm{bi}}=0,$ the ground state is a disordered state. The biorthogonal
order-disorder phase transition occurs at $J=\left \vert h\right \vert $ that
coincides the QPT from the gap closing. Fig.2(b) show the biorthogonal order
parameter from exact diagonal numerical calculation for 1D NH TI model with
$N=16$, in which the dotted red lines come from theoretical prediction.

\begin{figure}[ptb]
\includegraphics[clip,width=0.5\textwidth]{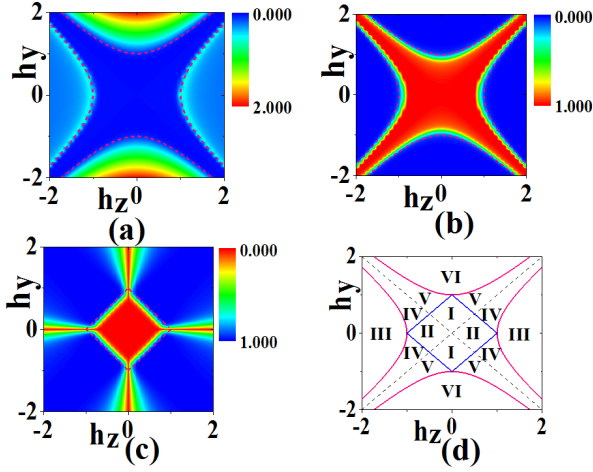}\caption{(Color online)
(a) The energy difference between the lowest two energy levels from exact
diagonal numerical calculations for 1D NH TI model with $N=16$; (b) The
biorthogonal order parameter from exact diagonal numerical calculation for 1D
NH TI model with $N=16$; (c) The state-similarity for the two degenerate
ground states $\left \vert \langle \mathrm{\overline{vac}}_{-}^{\mathrm{R}%
}(\beta)|\mathrm{\overline{vac}}_{+}^{\mathrm{R}}(\beta)\rangle \right \vert $
from exact diagonal numerical calculation for 1D NH TI model with $N=16$; (d)
The global phase diagram for theoretical predictions.}%
\end{figure}

Thirdly, we study the spontaneous \textrm{Z}$_{\mathit{2}}$-symmetry breaking
in this NH TI model.

We concentrate the DGSs in biorthogonal ordered phase with $A^{\mathrm{bi}}%
\neq0$. Because the order parameter changes sign under the biorthogonal
\textrm{Z}$_{\mathit{2}}$ transformation, i.e., $\left \langle
\mathrm{\overline{vac}}^{\mathrm{L}}(\beta)|\tilde{U}^{-1}(g^{\beta})\hat
{A}\tilde{U}(g^{\beta})|\mathrm{\overline{vac}}^{\mathrm{R}}(\beta
)\right \rangle =-A^{\mathrm{bi}}\neq A^{\mathrm{bi}}$, there must exist two
DGSs, $\left \vert \mathrm{\overline{vac}}_{+}(\beta)\right \rangle ,$
$\left \vert \mathrm{\overline{vac}}_{-}(\beta)\right \rangle .$ Under the
global biorthogonal \textrm{Z}$_{\mathit{2}}$-symmetric transformation, we have
$\tilde{U}(g^{\beta})\left \vert \mathrm{\overline{vac}}_{\pm}(\beta
)\right \rangle =\pm \left \vert \mathrm{\overline{vac}}_{\pm}(\beta
)\right \rangle $.

To show the NHSSB, we add a tiny longitudinal field on site $i_{0}$,
$\mathrm{\hat{H}}_{\mathrm{NTI}}^{\beta}\rightarrow(\mathrm{\hat{H}%
}_{\mathrm{NTI}}^{\beta})^{\prime}=\mathrm{\hat{H}}_{\mathrm{NTI}}^{\beta
}+\delta \mathrm{\hat{H}}$ where $\delta \mathrm{\hat{H}}=h^{x}\sigma_{i_{0}%
}^{x}$ with $h^{x}\ll J$.

To quantitatively demonstrate the NHSSB, we introduce the effective
Hamiltonian for the DGSs, $\mathcal{\hat{H}}_{\mathrm{GS}}=\left(
\begin{array}
[c]{cc}%
h_{++} & h_{+-}\\
h_{-+} & h_{--}%
\end{array}
\right)  $ where $h_{IJ}=\left \langle \mathrm{vac}_{I}^{\mathrm{R}}%
(\beta=0)\right \vert \mathrm{\hat{H}}_{\mathrm{NTI}}^{\beta}\left \vert
\mathrm{vac}_{J}^{\mathrm{R}}(\beta=0)\right \rangle ,$ $I,J=+,-$. $(%
\begin{array}
[c]{c}%
\left \vert \mathrm{vac}_{+}^{\mathrm{R}}(\beta=0)\right \rangle \\
\left \vert \mathrm{vac}_{-}^{\mathrm{R}}(\beta=0)\right \rangle
\end{array}
)$ are the basis of the DGSs for $\mathrm{\hat{H}}_{\mathrm{NTI}}^{\beta=0}$
under biorthogonal set. In general, we have $\left \vert \mathrm{\overline
{vac}}_{\pm}^{\mathrm{L/R}}(\beta)\right \rangle =c_{1,\pm}^{\mathrm{L/R}%
}\left \vert \mathrm{vac}_{+}^{\mathrm{L/R}}(\beta=0)\right \rangle +c_{2,\pm
}^{\mathrm{L/R}}\left \vert \mathrm{vac}_{-}^{\mathrm{L/R}}(\beta
=0)\right \rangle $ where $c_{1,\pm}^{\mathrm{L/R}}$ and $c_{2,\pm
}^{\mathrm{L/R}}$ are the complex parameters, respectively.

In the Hermitian limit $\beta \rightarrow0$, the effective Hamiltonian for the
DGSs $\mathcal{\hat{H}}_{\mathrm{GS}}$ is obtained as $\mathcal{\hat{H}%
}_{\mathrm{GS}}=\Delta \iota^{x}+\varepsilon \iota^{z}$ where $\Delta$\ is the
energy splitting from quantum tunneling effect that defined by $\Delta
=\left \langle \mathrm{vac}^{\mathrm{R}}|\mathrm{\hat{H}}_{\mathrm{NTI}}%
^{\beta}|\mathrm{vac}^{\mathrm{R}}\right \rangle =\frac{(h^{2}-J^{2})}%
{J}(-\frac{h}{J})^{N}$ and $\varepsilon=h^{x}$ is the energy difference
between the two DGSs. For the NH TI model, after considering a global ST on
the two DGSs $\mathcal{S}_{\mathrm{GS}}(\beta N)=(%
\begin{array}
[c]{cc}%
1 & 0\\
0 & e^{-\beta N}%
\end{array}
)$, the effective Hamiltonian for the DGSs is obtained as $\mathcal{\tilde{H}%
}_{\mathrm{GS}}^{(\beta N)}=\mathcal{S}_{\mathrm{GS}}(\beta N)\mathcal{\hat
{H}}_{\mathrm{GS}}\mathcal{S}_{\mathrm{GS}}^{-1}(\beta N)=\Delta^{+}\iota
^{+}+\Delta^{-}\iota^{-}+\varepsilon \iota^{z}$ where $\Delta^{+}=\Delta
e^{N\beta}=\frac{(h^{2}-J^{2})}{J}(-\frac{h}{J}e^{\beta})^{N}$ and $\Delta
^{-}=\Delta e^{-N\beta}=\frac{(h^{2}-J^{2})}{J}(-\frac{h}{J}e^{-\beta})^{N}.$
In thermodynamic limit $N\rightarrow \infty$, although $\Delta \rightarrow0,$
there exists the competition between the exponential decay of $\Delta$ with
the size of the system from quantum tunneling effect and the exponential
increase of $e^{\pm \beta N}$ with the size of the system from NH similarity
effect. Therefore, in thermodynamic limit there exist two phases: one phase is
$\left \vert \Delta e^{\pm \beta N}\right \vert \rightarrow0$, the other is
$\left \vert \Delta e^{\pm \beta N}\right \vert \rightarrow \infty$. At
$\left \vert \Delta e^{\pm N\beta}\right \vert =1$ (or $\pm \left \vert
h^{y}\right \vert \pm \left \vert h^{z}\right \vert =1$), the QPT occurs. The QPT
induced by perturbations at $\pm \left \vert h^{y}\right \vert \pm \left \vert
h^{z}\right \vert =1$ is accompanied by the sudden change of state-similarity
for the DGSs $\left \vert \langle \mathrm{\overline{vac}}_{-}(\beta
)|\mathrm{\overline{vac}}_{+}(\beta)\rangle \right \vert $. Fig.2(c) are the
numerical results for the state-similarity for the two DGSs. The dotted red
lines in Fig.2(c) is shown from the theoretical prediction, i.e.,
$\pm \left \vert h^{y}\right \vert \pm \left \vert h^{z}\right \vert =1$\cite{com2}.

On the one hand, in the region of $\left \vert \Delta e^{\pm N\beta}\right \vert
\rightarrow0$, the effective Hamiltonian for the DGSs is reduced into
$\mathcal{\tilde{H}}_{\mathrm{GS}}^{(\beta N)}\rightarrow \varepsilon \cdot
\iota^{z}$. The two DGSs are $\left \vert \mathrm{\overline{vac}}%
_{+}^{\mathrm{L/R}}(\beta)\right \rangle =\left \vert \mathrm{vac}%
_{+}^{\mathrm{L/R}}(\beta=0)\right \rangle $ and $\left \vert \mathrm{\overline
{vac}}_{-}^{\mathrm{L/R}}(\beta)\right \rangle =e^{-\beta N}\left \vert
\mathrm{vac}_{-}^{\mathrm{L/R}}(\beta=0)\right \rangle $. Now, the
state-similarity of the two DGSs is zero, i.e., $\left \vert \langle
\mathrm{\overline{vac}}_{-}(\beta)|\mathrm{\overline{vac}}_{+}(\beta
)\rangle \right \vert =0.$ In the thermodynamic limit $N\rightarrow \infty$, due
to the normalization factor for $\left \vert \mathrm{\overline{vac}}%
_{-}^{\mathrm{L/R}}(\beta)\right \rangle $ vanishes, i.e., $\left \vert
\langle \mathrm{\overline{vac}}_{-}^{\mathrm{R}}(\beta)|\mathrm{\overline{vac}%
}_{-}^{\mathrm{R}}(\beta)\rangle \right \vert =e^{-2\beta N}$, the quantum state
$\left \vert \mathrm{\overline{vac}}_{-}^{\mathrm{L/R}}(\beta)\right \rangle $
disappears and the ground state degeneracy $D$ becomes $1$. In particular, the
biorthogonal \textrm{Z}$_{\mathit{2}}$-symmetry are spontaneously broken
simultaneously, i.e., $\tilde{U}(g^{\beta})\left \vert \mathrm{\overline{vac}%
}^{\mathrm{R}}(\beta)\right \rangle \neq \left \vert \mathrm{\overline{vac}%
}^{\mathrm{R}}(\beta)\right \rangle $;\ On the other hand, in the region of
$\left \vert \Delta e^{\pm \beta N}\right \vert \rightarrow \infty$, the effective
Hamiltonian for the DGSs is reduced into $\mathcal{\tilde{H}}_{\mathrm{GS}%
}^{(\beta N)}\rightarrow \Delta e^{\pm \beta N}\iota^{+}$ or $\Delta e^{\pm \beta
N}\iota^{-}.$ The two DGSs are $\left \vert \mathrm{\overline{vac}}_{\pm
}^{\mathrm{L/R}}(\beta)\right \rangle =\frac{1}{\sqrt{2}}(\left \vert
\mathrm{vac}_{+}^{\mathrm{L/R}}(\beta=0)\right \rangle \pm e^{-\beta
N}\left \vert \mathrm{vac}_{-}^{\mathrm{L/R}}(\beta=0)\right \rangle )$. In this
region, the biorthogonal \textrm{Z}$_{\mathit{2}}$-symmetry are unbroken, i.e.,
$\tilde{U}(g^{\beta})\left \vert \mathrm{\overline{vac}}^{\mathrm{R}}%
(\beta)\right \rangle =\pm \left \vert \mathrm{\overline{vac}}^{\mathrm{R}}%
(\beta)\right \rangle .$ In the thermodynamic limit $N\rightarrow \infty$, the
state-similarity of the two DGSs is $1$, i.e., $\left \vert \langle
\mathrm{\overline{vac}}_{-}(\beta)|\mathrm{\overline{vac}}_{+}(\beta
)\rangle \right \vert =\left \vert \tanh(2\beta N)\right \vert \rightarrow1$ and
the ground degeneracy $D$ is also $1$. As a result, there is only one ground
state $\left \vert \mathrm{\overline{vac}}^{\mathrm{R}}(\beta)\right \rangle
=\left \vert \mathrm{vac}_{+}^{\mathrm{L/R}}(\beta=0)\right \rangle $ or
$\left \vert \mathrm{\overline{vac}}^{\mathrm{R}}(\beta)\right \rangle
=\left \vert \mathrm{vac}_{-}^{\mathrm{L/R}}(\beta=0)\right \rangle .$

In summary, in Fig.2(d), we plot the global phase diagram for the 1D NH TI
model: I is $\mathcal{PT}$ symmetric phase with BO and spontaneous biorthogonal
\textrm{Z}$_{\mathit{2}}$ symmetry breaking, II is $\mathcal{PT}$ symmetry
breaking phase with BO and spontaneous biorthogonal \textrm{Z}$_{\mathit{2}}$
symmetry breaking, III is $\mathcal{PT}$ symmetry breaking phase without BO
and with biorthogonal \textrm{Z}$_{\mathit{2}}$ symmetry, IV is $\mathcal{PT}$
symmetry breaking phase with BO and with biorthogonal \textrm{Z}$_{\mathit{2}}$
symmetry, V is $\mathcal{PT}$ symmetric phase with BO and with biorthogonal
\textrm{Z}$_{\mathit{2}}$ symmetry, VI is $\mathcal{PT}$ symmetric phase without BO and with biorthogonal \textrm{Z}%
$_{\mathit{2}}$ symmetry.

For 1D TI model with biorthogonal \textrm{Z}$_{\mathit{2}}$ symmetry, we can
use Jordan-Wigner transformation to map the original spin model $\hat
{H}_{\mathrm{NTI}}^{\beta}$ to a NH superconducting model and obtain the
correspondence exact results. That means the TI model with biorthogonal
\textrm{Z}$_{\mathit{2}}$ symmetry is an exactly solvable spin model.

We generalize the Jordan-Wigner transformation to the NH case by considering
global ST, i.e., $\tilde{c}_{j}=\mathcal{S}(\beta)c_{j}\mathcal{S}^{-1}%
(\beta)=\mathcal{S}(\beta)[(\prod_{k=1}^{j-1}\sigma_{k}^{y})(\sigma
_{j}^{z}-i\sigma_{j}^{x})]\mathcal{S}^{-1}(\beta)={\mathrm{e}^{-j\beta}}c_{j},$ and
$\tilde{c}_{j}^{\dagger}=\mathcal{S}(\beta)c_{j}^{\dagger}\mathcal{S}^{-1}%
(\beta)=\mathcal{S}^{-1}(\beta)[(\prod_{k=1}^{j-1}\sigma_{k}^{y})(\sigma_{j}%
^{z}-i\sigma_{j}^{x})]\mathcal{S}(\beta)={\mathrm{e}^{j\beta}}%
c_{j}^{\dagger}.$ From the Jordan-Wigner transformation, one can see
that the fermions get an additional "\emph{imaginary}" wave vector $k_{0}$,
i.e., $k_{0}=i\beta=i\ln(\frac{h^{y}+h^{z}}{h^{y}-h^{z}})$! The resulting
fermionic Hamiltonian corresponding to $\hat{H}_{\mathrm{NTI}}^{\beta}$
becomes $\hat{H}_{\mathrm{F}}^{\beta}=\sum_{k>0}\psi_{k}^{\dag}(H_{\mathrm{F}%
}^{\beta})\psi_{k}$ ($\psi_{k}^{\dag}=(c_{k}^{\dag},c_{k})$) where
\begin{equation}
H_{\mathrm{F}}^{\beta}=(-J\cos k+h^{y})\sigma^{z}+(-J\sin k+i(h^{z}))\sigma^{y}.
\label{4}%
\end{equation}
The fermion Hamiltonian $\hat{H}_{\mathrm{F}}^{\beta}$ can be "renormalized"
by an inverse ST $\mathcal{S}^{-1}(\beta)$ and becomes $\hat{H}_{\mathrm{F}%
}^{\beta=0}=\sum_{k>0}\tilde{\psi}_{k}^{\dag}(\tilde{H}_{\mathrm{F}}^{\beta
=0})\tilde{\psi}_{k}$ ($\tilde{\psi}_{k}^{\dag}=(\tilde{c}_{k}^{\dag}%
,\tilde{c}_{k})$) where
\begin{equation}
\tilde{H}_{\mathrm{F}}^{\beta=0}=(-J\cos k+h)\sigma^{z}+(-J\sin k)\sigma^{y}
\label{5}%
\end{equation}
(with $h=\sqrt{\left \vert h^{y}\right \vert ^{2}-\left \vert h^{z}\right \vert
^{2}}$). The detailed discussion about this issue is given in supplementary materials.

From Eq.(\ref{4}) and Eq.(\ref{4}), we obtain the same global phase diagram
for the 1D NH TI model as shown in Fig.2(d): The QPT for $\mathcal{PT}$
spontaneous symmetry breaking from Eq.(\ref{5}) also occurs at $\left \vert
h^{y}\right \vert =\left \vert h^{z}\right \vert $; The QPT corresponding to the
biorthogonal order-disorder phase transition at $J=\left \vert h\right \vert $ is
characterized by the gap closing for $\tilde{H}_{\mathrm{F}}^{\beta=0}$ from
Eq.(\ref{5}) (the purple lines in Fig.2(d)). In particular, biorthogonal
order-disorder phase transition at $\left \vert h\right \vert =J$\ in original
spin model corresponds to "topological phase transition" for the single body
fermion Hamiltonian $\tilde{H}_{\mathrm{F}}^{\beta=0}$. Based on the single
body fermion Hamiltonian $\tilde{H}_{\mathrm{F}}^{\beta=0},$ we define a
winding number$,$ $w=\frac{1}{2\pi}\int_{-\pi}^{\pi}\partial_{k}\phi(k)\cdot
dk$ where $\phi(k)=\tan^{-1}(d_{y}/d_{x})$ with $d_{y}=-J\sin k$ and
$d_{x}=h-J\cos k$. So, there are two phases: "topological phase" with $w=1$ in
the region of $\left \vert h\right \vert <J$ and trivial phase with $w=0$ in the
region of $\left \vert h\right \vert >J$.; The QPT corresponding to spontaneous
symmetry breaking at $\pm \left \vert h^{y}\right \vert \pm \left \vert
h^{z}\right \vert =1$ is characterized the gap (the gap for the real part of
energy levels) closing for $H_{\mathrm{F}}^{\beta}$ from Eq.(\ref{4}) (the
blue lines in Fig.2(d)).

\textbf{Conclusion and discussion:}\textit{ }In this paper, we develop the
theory for non-Hermitian spontaneous symmetry breaking. Universal features of
NH many-body systems with biorthogonal \textrm{Z}$_{\mathit{2}}$ symmetry are
explored. We find that the effect of NH terms splits the usual spontaneous
symmetry breaking (at $\lambda=\lambda_{s}$) from a biorthogonal
order-disorder phase transition (at $\lambda=\lambda_{c}$) with $\lambda
_{s}\neq \lambda_{c}$. As an exactly solvable spin model, we take the 1D NH
transverse Ising model $\mathrm{\hat{H}}_{\mathrm{NTI}}^{\beta}$ as example to
investigate the anomalous spontaneous symmetry breaking.

In addition, we also studied a two dimensional (2D) TI model with biorthogonal
\textrm{Z}$_{\mathit{2}}$-symmetry on square lattice and obtained a global
phase diagram that is quite similar to that of the 1D case (Fig.2(d)). And, for
1D or 2D TI\ model with biorthogonal \textrm{Z}$_{\mathit{2}}$-symmetry, the
NHSSB shows the same universal features. The detailed discussion are shown in
supplementary materials. In the future, we generalize the theory for
non-Hermitian spontaneous symmetry breaking to other models with discrete or
continuum non-unitary symmetries.

\acknowledgments This work is supported by NSFC Grant No. 11674026, 11974053, 11704186.


\begin{thebibliography}{99}
\bibitem {com}This Hamiltonian shows global $\mathcal{PT}$-symmetry, i.e.,
$\left[  \mathcal{P},\mathrm{\hat{H}}_{\mathrm{NTI}}^{\beta}\right]  \neq0$
and $\left[  \mathcal{T},\mathrm{\hat{H}}_{\mathrm{NTI}}^{\beta}\right]
\neq0$, but $\left[  \mathcal{PT},\mathrm{\hat{H}}_{\mathrm{NTI}}^{\beta
}\right]  =0.$ Here the time reversal operator $\mathcal{T}$ is defined as
$\mathcal{T}i\mathcal{T=-}i$ and the spin rotation operator $\mathcal{P}%
=\Pi_{j}(i\sigma_{j}^{z})$.
\bibitem {com1}For the case of $\left \vert h^{y}\right \vert <\left \vert
h^{z}\right \vert $, the energy spectra for the excitations become complex.
Now, we derive the QPT of gap closing at $J=\left \vert h\right \vert $ by
assuming that QPT occurs when the real part of the energy gap is closing.
\bibitem {com2}For the case of $\left \vert h^{y}\right \vert <\left \vert
h^{z}\right \vert $, the energy spectra for the excitations become complex.
Now, we derive the QPT of spontaneous biorthognal \textrm{Z}$_{\mathit{2}}%
$-symmetry breaking at $\pm \left \vert h^{y}\right \vert \pm \left \vert
h^{z}\right \vert =1$ by assuming that the ground states are the quantum states
with lowest real parts of energies.
\end{thebibliography}
\end{document}